\documentclass{article}
\usepackage{spconf}
\usepackage{amsmath,graphicx,hyperref,xcolor,amsfonts}
\usepackage{cite}
\usepackage{amsmath,amssymb,amsfonts}
\usepackage{algorithmic}
\usepackage{textcomp}
\usepackage{graphicx,color}
\usepackage{hyperref}

\hypersetup{
    colorlinks=true,
    citecolor=blue,
    linkcolor=blue,
    filecolor=magenta,      
    urlcolor=blue,
}
\usepackage{enumitem}
\setlist[1]{itemsep=1pt}

\usepackage{booktabs}
\graphicspath{{figures/}} 

\title{{ICASSP 2023} Deep Noise Suppression Challenge}
\name{
\parbox{\linewidth}{\centering
Harishchandra Dubey, Ashkan Aazami, Vishak Gopal, Babak Naderi, Sebastian Braun, Ross Cutler, \\
Alex Ju, Mehdi Zohourian, Min Tang, Hannes Gamper, Mehrsa Golestaneh, Robert Aichner}}
\address{Microsoft Corporation, Redmond, USA \\
    firstname.lastname@microsoft.com}
\begin{document}
\ninept
\maketitle
\begin{abstract}
Deep Speech Enhancement Challenge is the 5th edition of deep noise suppression (DNS) challenges organized at ICASSP 2023 Signal Processing Grand Challenges. DNS challenges were organized during 2019-2023 to stimulate research in deep speech enhancement (DSE)~\cite{dubey2022icasspdns}. Previous DNS challenges were organized at INTERSPEECH 2020, ICASSP 2021, INTERSPEECH 2021, and ICASSP 2022. From prior editions, we learnt that improving signal quality (SIG) is challenging particularly in presence of simultaneously active interfering talkers and noise. This challenge aims to develop models for joint denosing, dereverberation and suppression of interfering talkers. When primary talker wears a headphone, certain acoustic properties of their speech such as direct-to-reverberation (DRR), signal to noise ratio (SNR) etc. make it possible to suppress neighboring talkers even without enrollment data for primary talker. This motivated us to create two tracks for this challenge: (i) Track-1 Headset; (ii) Track-2 Speakerphone. Both tracks has fullband (48kHz) training data and testset, and each testclips has a corresponding enrollment data (10-30s duration) for primary talker. Each track invited submissions of personalized and non-personalized models all of which are evaluated through same subjective evaluation. Most models submitted to challenge were personalized models, same team is winner in both tracks where the best models has improvement of 0.145 and 0.141 in challenge's \textbf{Score} as compared to noisy blind testset.
\end{abstract}
%
%
\maketitle
%
\section{INTRODUCTION}
Deep Speech Enhancement (DSE) models perform significantly better than their classical counterparts~\cite{choi2020phase, reddy2020interspeechdns,dns2021interspeech,icassp2021challenge, dubey2022icasspdns,xia2020weighted}. We open-sourced training datasets and test sets, Personalized ITU-T P.835 subjective evaluation framework~\cite{naderi2021} as part of the previous DNS challenges. Our GitHub repository\footnote{~\url{https://github.com/microsoft/DNS-Challenge}} open-sourced Personalized and Non-personalized DNSMOS P.835~\cite{reddy2022dnsmos} and word accuracy (WAcc) APIs to empower iterative model improvements for teams articipating in the challenge. This reduces the barrier to entry in field and provides standard tools for evaluation of DSE models. Like previous challenges, each track has two testsets: (i) development (dev) test set which was released at the beginning of the challenge; (ii) blind test set released a few days before the final challenge deadline. While dev testset enables intermediate model evaluations, blind set is used for final ranking of models based on challenge metric (Score). Registration and submission of enhanced clips was done through CMT site~\url{https://cmt3.research.microsoft.com/DNSChallenge2023}, cloud storage and Azure Blobs as per participant's preference. Questions related to challenge can be sent to~\url{dns_challenge@microsoft.com}.

In this challenge, participants could use any datasets including external corpora, challenge training datasets to do model training. We do not prohibit the use of headset corpora for training speakerphone models and vice versa. Participants were required to describe the datasets used for training their models in sufficient detail in their extended journal papers, and provide a brief coverage in 2-page ICASSP grand challenge paper. Challenge website~\url{https://aka.ms/5th-dns-challenge} has details of scope and requirements; definitions of algorithmic latency, processing latency, causal model, Real-time factor (RTF) and associated challenge rules; and name of winning teams etc. We verified whether the top models are real-time or not and provide this information on our website. Verification was done with a $NRT$ testset which contains clips with overlapping segments. Previous challenge websites are linked at~\url{https://aka.ms/dns-challenge}.

 We introduced the following changes in this challenge: (i) there are two tracks: Headset, and Speakerphone both has desktop and mobile recordings in testsets; (ii) All testclips in both tracks has 10-30s enrollment speech (primary talker) with or without noise; (iii) Personalized P.835 framework is improved and now includes voice recognition, robust spam filtering and more accurate evaluation of enhanced testclips with noise and neighboring talkers; (iv) Personalized P.835 framework uses cleaned enrollment speech which was enhanced using a non-causal model. Cleaned enrollment speech helps human rater perform subjective evaluation more consistently.; (v) Personalized and non-personalized models for a track were sent to same subjective evaluation and ranked together, i.e. personalized and non-personalized models are treated alike and compared against each other. (vi) Track-1 and Track-2 has their separate subjective evaluations to ensure we have headset testclips evaluated together, and speakerphone testclips evaluated together.

 Most of the models submitted to challenge were personalized models. Winning teams in both tracks have similar ranking in each track. We invited challenge participants to submit their Preference-2 models which were not sent for subjective evaluation. These enhanced clips are expected to be used in semi-supervised training of personalized DNSMOS P.835. 
%
%
%
%
%
\section{Challenge Tracks}
%
Enhancing speech quality (i.e., eliminating speech distortion) and suppressing noise, reverberation and neighboring talkers are two trade-offs DSE models must handle. DSE models submitted to challenge were supposed to do joint denoising and dereverberation in presence of neighboring (interfering) talkers. The goal is to enhance audio signal and preserve the primary talker while suppressing the neighboring talkers, noise, and reverberation. All datasets used in this challenge were full band (48 kHz). It aimed to study headset and speakerphone DSE in separate tracks thereby creating possibility of new insights. Acoustic properties of headset scenarios can be leveraged in developing models for suppressing neighboring talkers without enrollment speech. Non-personalized model not requiring enrollment speech are suitable for consumers/industry with strict privacy requirements. 

This challenge has two tracks: Track-1 Headset; Track-2 Speakerphone. Dev and blind testset for both tracks were different. The testsets were collected using similar procedure except that the Track-1 testsets were collected using Headset devices while Track-2 testsets were collected using Speakerphone devices. Each test clip in both tracks has enrollment speech with 30s duration. The enrollment speech can be noise-free or noisy and with or without reverberation. This facilitates multi-condition enrollment of primary talkers which serves as a measure of robustness for personalized models which use enrollment speech as additional input for enhancing the testclips. Participants could choose to work on models with speaker enrollment or without it for one or both tracks. Each team was asked to submit 1-4 models depending on what models they trained. Each participating team could submit a maximum of one personalized and one non-personalized model for each track, e.g., a team can submit one personalized and one non-personalized model for Track 1 but not two personalized or two non-personalized model for Track 1. Similarly, another team could submit 4 models, personalized and non-personalized model for Track 1 and personalized and non-personalized model for Track 2. This rule facilitates almost equal representation of personalized and non-personalized models, and also similar participation in both tracks. 

While submitting the blind set enhanced clips, participants were asked to submit their preference for models submitted in each Track, e.g. they had to assign Preference-1 to their best model and Preference-2 to their second best model in each track. From the subjective evaluation of dev testsets, we noticed that subjective results were taking several days of time to get back the results. Also, very large batches of enhanced testclips may also results in some inconsistency in evaluation results. We used Preference-1 clips from all teams in both tracks for final evaluation based on enhanced blind set. Preference-2 enhanced clips could be leveraged in semi-supervised training of Personalized DNSMOS (PDNSMOS) P.835 model or in another subjective evaluation for generating labels for training of PDNSMOS P.835. All models for a track were evaluated and ranked together i.e., both personalized and non-personalized models for Track 1 went through one subjective evaluation. Similarly, for Track 2, all models in one subjective evaluation. Participants were encouraged to conduct experiments with both personalized and non-personalized models to elucidate the benefits of personalization. Though, this was not a requirement for this challenge. 

%
\section{Challenge Datasets}
\subsection{Training Data}
We used a machine learning model for extracting the near-field headset-like clean speech which is released as clean speech for headset track. We used entire clean speech from 4th DNS Challenge as clean speech for speakerphone track. Noise dataset and impulse responses are same as in 4th DNS Challenge~\cite{dubey2022icasspdns}. Participants can re-use near-field clean speech in Track 1 by convolving it with impulse responses for generating the training data for Track 2. The provided datasets include different languages with talker and device variety. We also provide speaker ID information for all clean speech clips to facilitate development of personalized models for both tracks. We also provide code for extracting speaker embeddings based on state-of-the-art ECAPA-TDNN embeddings trained on Voxceleb~\cite{dawalatabad2021ecapa, desplanques2020ecapa, model_ecapa}. Along with clean speech, we added clean speech with emotions such as crying, yelling, laughter, or singing to training data. Entire training set consists of clean speech clips in English and 10 non-English languages. 

Clean speech in the Track-2 training set is a total of 760.53 hours: read speech (562.72 hours), singing voice (8.80 hours), emotional speech (3.6 hours), Chinese Mandarin (185.41 hours). Track-1 clean speech is obtained by doing near-end speech extraction on Track-2 dataset. Clean speech has four subsets: (i) Read speech recorded in clean conditions; (ii) Singing clean speech; (iii) Emotional clean speech; and (iv) non-English clean speech. 

Noise data included in the training set is chosen from AudioSet~\cite{gemmeke2017audioset} and is identical to noise set in 4th DNS Challenge~\cite{dubey2022icasspdns}. AudioSet is a collection of about 2 million human-labeled 10s sound clips extracted from YouTube videos. There are over a million clips in AudioSet with audio classes music and speech and less than 200 clips for classes such as toothbrush, creak, etc. Approximately 42\% of the clips have a single class, but the rest may have 2 to 15 labels. Hence, we developed a sampling approach to balance the dataset in such a way that each class has at least 500 clips. We also used a speech activity detector to remove the clips with any kind of speech activity, to strictly separate speech and noise data. Audioset clips were made available at 44.1kHz, we upsampled those to fullband (48 kHz). The resulting noise dataset has 152 audio classes and 60,000 clips~\cite{dubey2022icasspdns}. In total, there are 181 hours of noise data in the training set. 
\subsection{Development Testset}
Both test sets consist of fullband audio clips recorded in real-world scenarios collected through crowd-sourcing where workers read provided text prompts and record their voice using desktop/laptop/mobile devices in the presence of noise and/or neighboring talkers. We include some new noise types in the test set covering relevant real-world scenarios, device variety and added paralinguistic test set as new category. Our dev test set consists of real-world test clips recorded by crowd-sourced workers.

The development test set for the non-personalized track consists of 600 real recordings. All clips contain noisy speech in the English language. Among these, 193 test clips have emotional speech in the presence of noise. There are six emotion types, namely happy, sad, angry, yelling, crying, and laughter. Crowd-sourced workers were asked to read provided text prompts and create emotional events in each test clip. The remaining clips contain the voice of a talker reading text in the presence of the following noise types: fan, air conditioner, typing, door shutting, clatter noise, car noise (i.e., standing near a car on a busy street or standing outside the car), kitchen noise (noise from kitchen utensils, dish scrubbing etc.), dish washer, running water, opening chips bags, munching or eating, creaking chair, heavy breathing, copy machine, baby crying, dog barking, inside-car noise (e.g., sitting on a passenger seat in a car which is being driven by someone else), mouse clicks, mouse scroll wheel, touch pad clicks, etc. Each test clip was recorded at 48 kHz with a duration of 10 to 20 seconds. Workers were asked to record in the near-field (close-talk) and far-field with distances of 1, 2, and 3 meters. All test clips in the non-personalized development test set were recorded using a laptop or desktop computer. Both development and blind test set have 2.5 minutes of enrollment speech for primary talkers to be used in personalized denoising. PDNS leverages speaker embedding (features) for preserving only the primary talker in a noisy environment while suppressing the neighboring talkers and noise. The development test set for the personalized track consists of 1443 real recordings. All clips contain noisy speech in the English language. Among these, 193 test clips have emotional speech in the presence of noise and are identical to the emotional test clips in the non-personalized track. There are 737 test clips where the primary talker reads the provided text in the presence of the same noise types as those in the non-personalized track. Each test clip was recorded at 48~kHz with a duration of 10--20 seconds. Workers were asked to record in the near-field (close-talk) and far-field with distances of 1, 2, and 3 meters. There are 166 test clips with the primary talker speaking in the presence of a neighboring talker and noise where both the noise and neighboring talker are simultaneously active in the primary talker's background. There are 347 test clips where the primary talker is speaking in the presence of a neighboring speaker with no background noise. Thus, we have simulated three scenarios for PDNS: (i) primary talker in the presence of noise; (ii) primary talker in the presence of neighboring talker; and (iii) primary talker in the presence of simultaneously active neighboring talker and noise. All test clips in the personalized development test set were recorded using a laptop or desktop computer.
\subsection{Blind Testset}
We include some new noise types in the test set covering relevant real-world scenarios, device variety and added paralinguistic test set as new category. Blind testset consists of real testclips which were not previously used in any challenge and not otherwise available publically. Our test set consists of real-world test clips recorded by crowd-sourced workers. Some of the devices used for collecting blind set includes [list of devices]. Blind set clips were chosen to confirm blind set specifications which were also shared with crowd-sourced workers. We performed rigorous quality assurance (QA) to ensure blind set is representative of real-world scenarios in terms of speaker variety, device variety, variety in acoustic scenarios, different direct-to-reverb-ratio (DRR), different T60 which is achieved by changing the relative and absolute position of primary and interfering talkers, noise source and presence of reflecting surfaces etc.

Unlike previous challenges, Blind testset in this challenge contains paralinguistic test clips. These contains standard forms of paralanguage~\cite{paralanguage} including but not limited to The throat-clear, "hmm" or "mhm", "Huh?" or  "what?, Gasps, Sighs, Moans and groans, Deceptive speech, Sincere speech, Speech with high-base, Speech with high-pitch, Speech with low-pitch, Confident speech, Tired speech (when talker is tired), Persuasive speech, Voice change mid-clip (i.e. mimicry in last 50\% of the clip). 

Blind testset will also include emotional speech including but not limited to happy, sad, angry, yelling, crying, and laughter. Blind testset include real test clips with high reverberation, high reverberation with noise, and noise in presence of interfering talkers. Testset noises include but not limited to Office scenarios (Typing, AC, Door shutting, Eating/munching, Copy machine, Squeaking chair, Notification sounds etc.), Home scenarios (Baby crying, Dogs, TV, Radiators, hair dryer, kitchen noise, running water etc.), Appliances (Washer Dryer, Dishwasher, Coffee maker, kitchen noise, Vacuum cleaner etc.), Fire alarm, Car, Inside parked car on busy road, In-car neighboring talkers, Traffic Road, Car noise (from machinery, control systems, turn signal etc.), Café, Coffee machine, Blender, Background babble, Airport Announcements etc. 
%
\begin{figure*}[!t]
\includegraphics[width=0.98\textwidth]{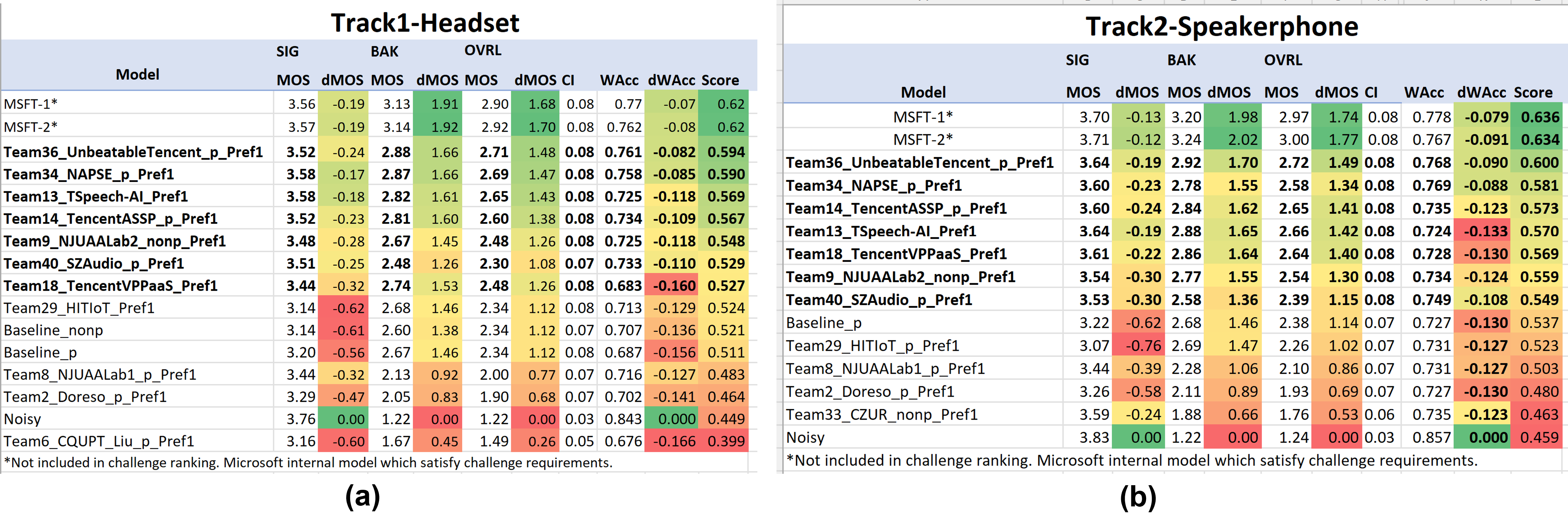}
\caption{Results: Personalized P.835 subjective evaluation, Word Accuracy (WAcc) and Challenge metric (Score) computed on blind testset for all teams in (a) Track-1 Headset; (b) Track-2 Speakerphone.}
\label{fig:blind_results}
\end{figure*}
\section{Evaluation Setup}
\subsection{Baseline Models}
Along with training datasets and testsets, we also provide a baseline model (or enhanced clips) for both tracks. 

Baseline models were personalized and non-personalized variants of models presented in~\cite{eskimez2022real}.
\subsection{Subjective Evaluation}
\subsubsection{Cleaning Enrollment Clips}
We cleaned (enhanced) enrollment clips with a non-causal model based on E3Net architecture~\cite{thakker2022fast}. The subjective evaluation uses only 5s of enrollment speech. We manually choose a 5s segment either from the enhanced enrollment clip ensure all long pause (>0.2s) are removed and resulting 5s audio is normalized to ensure easier recognition of primary talker by human raters. The non-causal DSE model is a fullband time domain domain based on end-to-end enhancement network (E3Net) which was originally proposed for the task of personalized speech enhancement using wideband signals~\cite{thakker2022fast}. 

Non-causal model is non-personalized and is trained to remove background noise and reverberation for fullband speech. This model uses a learnable encoder and decoder instead of STFT and iSTFT which can mitigate the problem of imperfect phase reconstruction that exists in most time-frequency based speech enhancement methods. This model is based on 20ms frames with a stride of 10ms. The core block in non-causal model is a bidirectional LSTM which offers improved performance for offline processing which is suitable for enhancing enrollment clips for subjective evaluations. 

This model was trained on training clean speech and noise from he 4th DNS challenge~\cite{dubey_icassp_2022}. Impulse responses used for training this model consists of 150,000 simulated impulse responses at sampling rate 48kHz using the Image Source method~\cite{allen1979image}. The reverberant speech and noise signals were mixed with a signal-to-noise ratio (SNR) drawn from a Gaussian distribution with N(-5,20) dB. All training samples were 10s segments. Training data consists of synthesized 1000 hours of fullband data. For increasing the robustness of the model with respect to audios with different bandwidths, namely narrowband and wideband audios, we also downsampled half of the 1000 hours training set to 16kHz and 8kHz respectively and upsampled them back to 48 kHz and combined these with the training set. Thus, the final training dataset has 3000 hours of data. 
\subsubsection{Personalized P.835 Framework}
This challenge relies on ITU-T P.835~\cite{naderi2021} subjective evaluation framework. A modified version (Personalized) ITU-T P.835 was used for measuring the performance of personalized DSE models. Personalized P.835 framework uses 5s of clean enrollment speech for primary talkers to help human raters recognize the primary talker's voice while assigning subjective scores. Human raters were instructed to focus on the quality of the voice of the primary talker when two or more talkers were present in a test clip. In addition, personalized P.835 subjective framework is improved to include voice recognition, better spam filtering and more accurate evaluation of enhanced testclips with noise and neighboring talkers. 

In all test clips, we have only one primary talker in enrollment clip while noisy test clip may have noise, reverberation, one or more neighboring talkers in addition to a primary talker. The goal of DSE models is to preserve primary talker's speech while suppressing everything else. 

Crowd-sourced workers doing subjective evaluation are instructed to rate interfering talkers as undesirable signal so the model which suppresses interfering talkers is rated higher. Personalized P.835 subjective framework provides three scores, namely speech quality (SIG), background noise quality (BAK), and overall audio quality (OVRL). We evaluated all challenge models through our Personalized P.835 framework to get the subjective ratings used for computing the challenge metric. 
\subsection{Word Accuracy}
We computed word accuracy (WAcc) from a state-of-the-art speech recognition system. We did WAcc computation ourselves during last week of the challenge to ensure all models are evaluated in exactly same way. WAcc is an objective metric for measuring the impact of speech enhancement on speech recognition transcription service. WAcc is defined as \\ $\text{WAcc} = 1 - \text{WER}$,

where WER is the word error rate of speech recognition system. We transcribed the entire blind set for both tracks to obtained the ground-truth. Dev testset was not transcribed. We computed WAcc results internally. Unlike subjective P.835 framework which uses only 7s of manually chosen segment from noisy or enhanced clips, WAcc engine uses entire testclips. Blind set test clips are of different duration from 10s to more than 6 minutes. 

Blind testset was collected from crowd-sourced data collection using various recording apps where the worker was provided with a text prompts to read from. Since there are often reading errors, omissions of words etc., prompts do not reflect the correction transcription. We used a five step approach to obtain ground-truth transcription for blind testset. In first step, we just obtain the prompts for each testclips in blind set. In second step, we obtained the transcripts from the state-of-the-art speech recognition engine for each testclips in blind set. In third step, expert human listeners listened each testclips and generated the corresponding human transcripts. Expert listeners were advised to listen audio clips several times until they were confident about their transcription. In fourth step, we compute word error rate (WER) for each testclip in blind set. Then, the testclips with WER > 0.5 were chosen for fifth round of listening. In fifth round of listening, human listeners listen the clips with WER > 0.5 and validated/corrected the human transcriptions. We found that a very few clips were needed to be corrected at fifth stage this elucidates the robustness of our transcription approach. 

We found that several clips in blind set results had WER > 0.5, it is often due to insertion error where DSE models leaked the secondary talker's voice and those speech were transcribed by speech recognition engine. During five step approach for generating ground-truth transcription, we found out that there were several clips where neighboring talker was sounding very similar to primary talker (e.g., same accent, same gender, similar loudness and sitting close to each other), some of those clips took more than 10 times listening to correctly transcribe, some of those were not possible to transcribe and hence skipped. Thus, our WAcc was computed over entire duration of selected clips with ground-truth transcripts. It is important to note that speech recognition engine is not personalized and does not use any enrollment speech to focus on primary talker. In future, it would be interesting to develop a personalized speech recognition engine. WAcc ground-truth transcripts only contained words spoken by the primary talker, thus treating interfering talker as undesirable signal. 
\subsection{Challenge Metric}
Like past challenges, models in both tracks were ranked in terms of a final score obtained by weighted average of subjective P.835 scores and word accuracy (WAcc). Four metrics, three personalized P.835 subjective scores namely SIG, BAK, OVRL and WAcc from a speech recognition system were used to evaluate challenge models. Metrics on the blind test set were combined into a final score for ranking the models. Higher WAcc shows superior speech enhancement performance with respect to speech recognition. Higher P.835 scores shows better subjective speech quality. The final score is computed as \\ $\text{Final score}= 0.5[\text{WAcc} + 0.25(\text{OVRL} -1)]$. 

We evaluated the submitted models based on Personalized ITU-T P.835 subjective evaluation scores, namely speech quality (SIG), background noise quality (BAK), and overall audio quality (OVRL), and WAcc from a state-of-the-art speech recognition system. WAcc is an objective metric for measuring the impact of speech enhancement on speech recognition transcription service.
\section{Results~\& Discussions}
There were 11 submissions in Track-1 and 11 submissions in Track-2 out of which 10 teams participated in both tracks. Almost all models in both tracks were personalized models. We had two baseline models for Track-1 out of which one was non-personalized model. We had personalized model as baseline for Track-2. By dint of acoustic properties of primary speech in a headset scenarios, a non-personalized model may be able to suppress neighboring talkers and noise, hence we included a non-personalized model. Each team was asked to provide enhanced clips from their models. Teams were allowed to submit maximum of two models (at most one personalized and at most one non-personalized) per track. Since subjective evaluations may take very long-time if number of submitted models is large, we ask participants to provide their Preference of models. We chose the first Preference model from each team in their respective tracks for subjective evaluations. We also conducted a dev testset subjective evaluation based on enhanced dev set and provided the results to participants.  

Fig.~\ref{fig:blind_results} show the subjective personalized P.835 scores, WAcc and challenge metric (Score) for all teams sorted in decreasing order of performance. dMOS for SIG, BAK, OVRL refers to the difference in SIG, BAK, OVRL between the enhanced clip and corresponding noisy clip. Similarly, dWAcc is the difference in WAcc between the enhanced clip and noisy clip. We conducted ANOVA test on top models in each track to determine statistical significance (see~\url{https://aka.ms/5th-dns-challenge}) for ANOVA results. 
%
\section{Conclusions}
This challenge has a more diverse blind test set collected through crowdsourcing using multiple data vendors. We included paralinguistic testclips and leakage testclips in blind set. We enhanced and chose the enrollment speech segment for primary talkers, and noisy segment from testclips to ensure robust subjective evaluations. We verified if winning models are causal with helps of our $NRT$ testset. Results show degradation in signal quality (SIG) for most of the models. It may be due to two reasons: (i) testset is significantly challenging; (ii) Personalized model end up suppressing the primary talker. Noticeable SIG degradation's happen due to suppression of primary talker’s speech and/or leakage of interfering talker and noise. Detailed analysis of results from current and previous DNS challenges will be covered in our extended OJSP journal paper. We encourage all the challenge participants and readers to send us their feedback and contribute to DNS Challenge repository.
\bibliographystyle{IEEEbib}
\bibliography{main}

\begin{thebibliography}{10}

\bibitem{dubey2022icasspdns}
Harishchandra Dubey, Vishak Gopal, Ross Cutler, Ashkan Aazami, Sergiy
  Matusevych, Sebastian Braun, Sefik~Emre Eskimez, Manthan Thakker, Takuya
  Yoshioka, Hannes Gamper, et~al.,
\newblock ``{ICASSP 2022} deep noise suppression challenge,''
\newblock in {\em ICASSP 2022-2022 IEEE International Conference on Acoustics,
  Speech and Signal Processing (ICASSP)}. IEEE, 2022, pp. 9271--9275.

\bibitem{choi2020phase}
Hyeong-Seok Choi, Hoon Heo, Jie~Hwan Lee, and Kyogu Lee,
\newblock ``Phase-aware single-stage speech denoising and dereverberation with
  {U}-net,''
\newblock {\em arXiv preprint arXiv:2006.00687}, 2020.

\bibitem{reddy2020interspeechdns}
Chandan~KA Reddy, Vishak Gopal, Ross Cutler, Ebrahim Beyrami, Roger Cheng,
  Harishchandra Dubey, Sergiy Matusevych, Robert Aichner, Ashkan Aazami,
  Sebastian Braun, et~al.,
\newblock ``The interspeech 2020 deep noise suppression challenge: Datasets,
  subjective testing framework, and challenge results,''
\newblock {\em arXiv preprint arXiv:2005.13981}, 2020.

\bibitem{dns2021interspeech}
Chandan~KA Reddy, Harishchandra Dubey, Kazuhito Koishida, Arun Nair, Vishak
  Gopal, Ross Cutler, Sebastian Braun, Hannes Gamper, Robert Aichner, and
  Sriram Srinivasan,
\newblock ``{INTERSPEECH 2021 Deep Noise Suppression Challenge},''
\newblock {\em ISCA INTERSPEECH}, 2021.

\bibitem{icassp2021challenge}
Chandan K.~A. Reddy, Harishchandra Dubey, Vishak Gopal, Ross Cutler, Sebastian
  Braun, Hannes Gamper, Robert Aichner, and Sriram Srinivasan,
\newblock ``{ICASSP 2021 Deep Noise Suppression Challenge},''
\newblock in {\em IEEE ICASSP}, 2021, pp. 6623--6627.

\bibitem{xia2020weighted}
Yangyang Xia, Sebastian Braun, Chandan~KA Reddy, Harishchandra Dubey, Ross
  Cutler, and Ivan Tashev,
\newblock ``Weighted speech distortion losses for neural-network-based
  real-time speech enhancement,''
\newblock in {\em ICASSP 2020-2020 IEEE International Conference on Acoustics,
  Speech and Signal Processing (ICASSP)}. IEEE, 2020, pp. 871--875.

\bibitem{naderi2021}
Babak Naderi and Ross Cutler,
\newblock ``Subjective evaluation of noise suppression algorithms in
  crowdsourcing,''
\newblock in {\em ISCA INTERSPEECH}, 2021.

\bibitem{reddy2022dnsmos}
Chandan~KA Reddy, Vishak Gopal, and Ross Cutler,
\newblock ``Dnsmos p. 835: A non-intrusive perceptual objective speech quality
  metric to evaluate noise suppressors,''
\newblock in {\em ICASSP 2022-2022 IEEE International Conference on Acoustics,
  Speech and Signal Processing (ICASSP)}. IEEE, 2022, pp. 886--890.

\bibitem{dawalatabad2021ecapa}
Nauman Dawalatabad, Mirco Ravanelli, Fran{\c{c}}ois Grondin, Jenthe Thienpondt,
  Brecht Desplanques, and Hwidong Na,
\newblock ``Ecapa-tdnn embeddings for speaker diarization,''
\newblock {\em arXiv preprint arXiv:2104.01466}, 2021.

\bibitem{desplanques2020ecapa}
Brecht Desplanques, Jenthe Thienpondt, and Kris Demuynck,
\newblock ``Ecapa-tdnn: Emphasized channel attention, propagation and
  aggregation in tdnn based speaker verification,''
\newblock {\em arXiv preprint arXiv:2005.07143}, 2020.

\bibitem{model_ecapa}
``{DNSMOS Git Repo},''
  \url{https://huggingface.co/speechbrain/spkrec-ecapa-voxceleb},
\newblock [Online; accessed 2022-09-01].

\bibitem{gemmeke2017audioset}
Jort~F Gemmeke, Daniel~PW Ellis, Dylan Freedman, Aren Jansen, Wade Lawrence,
  R~Channing Moore, Manoj Plakal, and Marvin Ritter,
\newblock ``Audio set: An ontology and human-labeled dataset for audio
  events,''
\newblock in {\em 2017 IEEE international conference on acoustics, speech and
  signal processing (ICASSP)}. IEEE, 2017, pp. 776--780.

\bibitem{paralanguage}
``{Paralanguage},'' \url{https://en.wikipedia.org/wiki/Paralanguage},
\newblock [Online; accessed 2022-09-01].

\bibitem{eskimez2022real}
Sefik~Emre Eskimez, Takuya Yoshioka, Alex Ju, Min Tang, Tanel Parnamaa, and
  Huaming Wang,
\newblock ``Real-time joint personalized speech enhancement and acoustic echo
  cancellation with e3net,''
\newblock {\em arXiv preprint arXiv:2211.02773}, 2022.

\bibitem{thakker2022fast}
Manthan Thakker, Sefik~Emre Eskimez, Takuya Yoshioka, and Huaming Wang,
\newblock ``Fast real-time personalized speech enhancement: End-to-end
  enhancement network (e3net) and knowledge distillation,''
\newblock {\em arXiv preprint arXiv:2204.00771}, 2022.

\bibitem{dubey_icassp_2022}
Harishchandra Dubey, Vishak Gopal, Ross Cutler, Ashkan Aazami, Sergiy
  Matusevych, Sebastian Braun, Sefik~Emre Eskimez, Manthan Thakker, Takuya
  Yoshioka, Hannes Gamper, and Robert Aichner,
\newblock ``{ICASSP} 2022 {Deep} {Noise} {Suppression} {Challenge},''
\newblock in {\em {ICASSP}}, 2022.

\bibitem{allen1979image}
Jont~B Allen and David~A Berkley,
\newblock ``Image method for efficiently simulating small-room acoustics,''
\newblock {\em The Journal of the Acoustical Society of America}, vol. 65, no.
  4, pp. 943--950, 1979.

\end{thebibliography}
\end{document}